\documentclass[longbibliography,superscriptaddress,showpacs,aps,twocolumn,prb,floatfix]{revtex4-1}
\usepackage{amssymb}
\usepackage{amsmath}
\usepackage{graphicx}
\usepackage{bm}

\usepackage[caption=false]{subfig}

\begin{document}
\title{Thermoelectric properties of a double quantum dot out of equilibrium in Kondo and intermediate valence regimes}

\author{D. Perez Daroca}
\affiliation{Gerencia de Investigaci\'on y Aplicaciones, GAIDI, CNEA, 1650 San Mart\'{\i}n, Buenos Aires, Argentina}
\affiliation{Consejo Nacional de Investigaciones Científicas y Técnicas, 1025 CABA, Argentina}

\author{P. Roura-Bas}
\affiliation{Centro At\'{o}mico Bariloche, GAIDI,  8400 Bariloche, Argentina}
\affiliation{Consejo Nacional de Investigaciones Científicas y Técnicas, 1025 CABA, Argentina}

\author{A. A. Aligia}
\affiliation{Instituto de Nanociencia y Nanotecnolog\'{\i}a 
CNEA-CONICET, GAIDI,
Centro At\'{o}mico Bariloche and Instituto Balseiro, 8400 Bariloche, Argentina}

\begin{abstract}
We study a system composed of two quantum dots connected in series between two leads at different temperatures, in the limit of large intratomic repulsion. Using the non-crossing approximation, we calculate the spectral densities at both 
dots $\rho_i(\omega)$,
the thermal and thermoelectric responses, thermopower and 
figure of merit  in different regimes. The interatomic repulsion
leads to finite heat transport even if the hopping between
the dots $t=0$. The thermopower can be very large compared to single-dot systems in several regimes. The changes in sign
of the thermoelectric current can be understood from the position and magnitude of the Kondo and charge-transfer peaks in 
$\rho_i(\omega)$. 
The figure of merit can reach values near 0.7.
The violation of the  Wiedemann-Franz  law is much more significant than in previously studied nanoscopic systems.
An analysis of the widths of $\rho_i(\omega)$ indicates that 
the dots are at effective temperatures $T_i$ intermediate between those of the two leads, which tend to be the same for large $t$.

\end{abstract}

%\pacs{73.23.-b, 71.10.Hf, 75.20.Hr}

\maketitle

\section{Introduction}

\label{intro}

During the last two decades, the study of nanodevices that convert heat into work has received great 
attention due to possible applications \cite{bene}.
In particular, thermoelectric properties in transport through single molecules
\cite{guo13,kim15,rincon,cui17,miao18,cui19} and quantum dots \cite{erdman,svilans,josef,dutta} 
has been experimentally studied. Usually, these systems can be modeled by the single impurity Anderson model, or its generalizations to multilevel systems, and the thermal 
properties of these models, in and out of equilibrium, have been addressed by several calculations \cite{cui17,erdman,boese,hump02,kra07,Kubala,pola,costi,leij,azem,see,ng,azem2,dorda16,sierra17,li17,li18,burkle,asym,KK,tesser,mana,mina,cortes}. 
Particularly in recent years, these studies have been extended to systems of two quantum dots or molecules \cite{tesser,donsa,craven,sierra,dare,yada,heat,lava20,lomba,zhang,ghosh,mana2}.

For linear response (in the limit of vanishing voltage
and thermal gradient) the numerical renormalization group (NRG)
has been used to calculate the relevant quantities, like the electrical  and thermal conductances or the thermopower 
(Seebeck coefficient $S$) \cite{dare,donsa,costi,anders08,Ngh18,mana,mina,mana2}.
The NRG is a robust numerical technique. However, out of equilibrium, for a finite bias voltage or
difference of temperature between the conducting leads, 
NRG is very difficult to apply (although some
developments using the scattering states NRG approach 
look promising \cite{anders08,Ngh18}) and different approximations have been 
used to capture the essential physics of the Kondo effect
\cite{hewson}, which is always present in nanostructures
for large on-site repulsion $U$ and degenerate configurations of the localized electrons 
(like odd number of electrons in one quantum dot). 
In its simplest and more usual realization, 
the Kondo effect can be described as the screening of a
localized spin by the surrounding free conduction electrons 
forming a many-body singlet.

Among the different approximations widely used to treat Kondo
systems out of equilibrium, one can mention equations of motion (EOM) 
for the Keldysh Green functions \cite{pola,li18,rome09,rapha,rome,crep}, renormalized perturbation 
theory (RPT) in $U$ or similar Fermi-liquid approaches \cite{ng,asym,KK,heat,oguri01,oguhe,mora09,sela,ct,karki18,tera20,tera}, slave bosons in the 
mean-field approximation (SBMFA) \cite{sierra17,sierra,geme,dong02,hamad13,aguado,rosa02}, non-crossing approximation 
(NCA) \cite{lomba,win,hettler,sitri1,sitri2,serge,tetta,vibra,desint19,choi,roura10},
auxiliary master equation approach \cite{Fugger18,Fugger20},
or functional renormalization-group,
restricted to large bias voltages or magnetic fields \cite{rosch03}. Recently, a method that combines NRG with
time-dependent density matrix renormalization has been used \cite{mana22}.
All traditional
methods have limitations. The EOM does not reproduce correctly the functional dependence of the energy scale 
$T_{K}$ on the on-site energy $E_{d}$ \cite{rapha,rome}.
RPT is valid for small energies and it is not easy
to apply to more complex systems. The SBMFA with increasing
temperature \cite{hewson} or magnetic field \cite{lady} has an abrupt artificial transition to
a phase in which the impurity decouples from the 
conduction band. The NCA does not satisfy 
Fermi-liquid relations at zero temperature, 
and when the ground-state configuration of the isolated
impurity (or system disconnected to the conducting leads)
is non degenerate (like for positive $E_d$ for one impurity
or non-zero magnetic field), the impurity self energy has an
unphysical positive imaginary part at low temperatures 
and as a consequence the spectral density  presents a spurious
peak at the Fermi energy \cite{win}.  
For finite $U$ the NCA ceases to reproduce correctly the dependence of $T_K$ with parameters 
and vertex corrections
should be included \cite{pruschke89,haule01,tosi11}. 

Using the SBMFA, Sierra \textit{et al.} 
studied a system of two Kondo impurities
(modeling two quantum dots in a serial arrangement)
under a thermal bias \cite{sierra}. 
They find some unusual results, like negative differential
thermal conductance and sign reversal of the 
thermoelectric current for certain parameters.
Taking into account the limitations mentioned above for
the SBMFA at finite temperature, it is convenient to check 
these results with an independent technique.
We study the same system (including also interdot Coulomb repulsion) using the NCA. In spite of the shortcomings mentioned above, the NCA is known to reproduce correctly the relevant energy scale $T_{K}$
and its dependence on the different parameters.
It has proved to be a
very valuable tool for calculating the differential electrical conductance
through  different systems \cite{hettler}
such as two-level quantum dots and  C$_{60}$ molecules 
displaying a quantum phase transition \cite{sitri1,sitri2,serge}, a nanoscale Si transistor \cite{tetta}, or vibrating molecules \cite{vibra,desint19}, among others \cite{lomba,choi}. It also reproduces correctly the scaling
of the conductance for small bias voltage $V$ and temperature $T$ \cite{roura10}.

Regarding the accuracy of the NCA within the linear response regime, we note that a comparative study of the numerical renormalization group and NCA results
for the spectral functions has been done \cite{costi96}. In particular, they have shown (Fig. 7 of Ref. \onlinecite{costi96}) that the NCA spectral functions at temperatures $T\ge 0.1T_K$ are reliable enough to compute transport properties. As for the non-equilibrium results at equal chemical potentials, the same is true if the temperature of the cold lead is kept at a similar condition, that is 
$T_R \ge 0.1 T_K$.

Our paper is a comprehensive study of different regimes
of the double quantum dot (DQD) system, although restricted to the situation with inversion symmetry for simplicity. Among our most peculiar results, we find unusually large and small Lorenz numbers 
compared to those found usually in  metals, semi-metals, alloys, degenerate semiconductors
\cite{kumar93} and other nanoscopic systems \cite{Kubala,burkle}. Also in some cases large figure of merits
$ZT$ are obtained and in other cases, large thermopower.
Analyzing the width of the spectral densities of both dots, we show that they can be interpreted
as each dot is at an intermediate temperature between that 
of the left and the right leads. For interdot hopping 
$t \rightarrow 0$ each dot is at equilibrium with the corresponding lead, while at large $t$ they are at 
nearly the same intermediate temperature.

The paper is organized as follows. In Sec. \ref{model}, the model, the methods and the expressions for the electrical and heat current are presented. The results for the occupancy of the DQD fluctuating  
between 0 and 1 particles and between 1 and 2 particles are showed in Secs. \ref{res01} and \ref{res12}, respectively. Sec. \ref{sum} is devoted to the summary and discussion. The NCA treatment of fluctuations between 1 and 2 particles in the DQD is discussed in Appendix \ref{anca}. Finally, in Appendix \ref{current-delta}, an alternative expression for the calculation of the charge current is described in detail.

\section{Model and methods}

\label{model}

\begin{figure}[h!]
\centering
\includegraphics[width=0.45\textwidth]{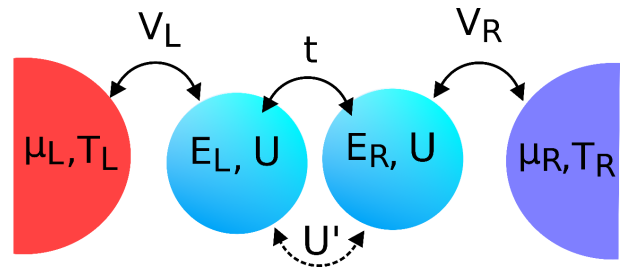}  
\caption{(Color online) Scheme of the system}
\label{scheme}
\end{figure}

A scheme of the system out of equilibrium is shown in Fig. \ref{scheme}. It is composed by a DQD where each dot
is tunnel-coupled to a metallic reservoir. The Hamiltonian describing the
system is the following

\begin{equation}
H=H_{DQD}+H_{c}+H_{V}.  \label{ham}
\end{equation}

The first term describes the DQD, 
\begin{eqnarray}
H_{DQD} &=&\sum_{\nu \sigma }E_{\nu }d_{\nu \sigma }^{\dagger }d_{\nu \sigma
}+\sum_{\nu }Un_{\nu \uparrow }n_{\nu \downarrow }  \notag \\
&&+U^{\prime }\sum_{\sigma \sigma ^{\prime }}n_{L\sigma }n_{R\sigma ^{\prime
}}\\
	&&-t\sum_{\sigma }\left( d_{L\sigma }^{\dagger }d_{R\sigma }
+\text{H.c.}\right) \notag  \label{hdqd}
\end{eqnarray}
where $E_{\nu }$ and $U$ ($U^{\prime }$) are the energy levels and the intra-
(inter-) dot Coulomb repulsion respectively,  $\nu =\{L,R\}$ labels the left
and right dot (and leads) and $\sigma ={\uparrow ,\downarrow}$ stands for
the spin projection. The hopping energy between dots is represented by $t$.

The second term describes the conducting leads 
\begin{equation}
H_{c}=\sum_{\nu}H_{\nu}=\sum_{k\nu \sigma }\varepsilon _{\nu k\sigma }\,c_{\nu k\sigma
}^{\dagger }c_{\nu k\sigma },  \label{hleads}
\end{equation}
and the last one, describes the hybridization between each dot and its
respective lead 
\begin{equation}
H_{V}=\sum_{k\nu \sigma }\left( V_{k\nu }\,d_{\nu \sigma }^{\dagger }c_{\nu
k\sigma }+\text{H.c.}\right) .  \label{hyb}
\end{equation}

In order to treat the system within the NCA and avoid the need to consider
vertex corrections, we diagonalize $H_{DQD}$ and retain only two neighboring
configurations, which correspond to the limit $U\rightarrow +\infty $. In
this way, the problem is mapped into a multilevel system connected to two
conducting leads. With some adequate change of phases, the model is
electron-hole symmetric. Therefore there are two independent cases to be
considered: fluctuations between 0 and 1 particles in the DQD and
fluctuations between 1 and 2 particles. Thus, It suffices to retain the
eigenstates and energies of $H_{DQD}$ \ for 1 and 2 particles. 

We assume for simplicity that the DQD has inversion symmetry. Therefore 
$E_{L}=E_{R}\equiv E_{d}$. The eigenstates for 1 particle are the even and
odd linear combinations

\begin{eqnarray}
d_{e\sigma }^{\dagger } &=&\frac{1}{\sqrt{2}}\left( d_{L\sigma }^{\dagger
}+d_{R\sigma }^{\dagger }\right) ,\text{ }E_{e}=E_{d}-t,  \notag \\
d_{o\sigma }^{\dagger } &=&\frac{1}{\sqrt{2}}\left( d_{L\sigma }^{\dagger
}-d_{R\sigma }^{\dagger }\right) ,\text{ }E_{o}=E_{d}+t,  \label{deo}
\end{eqnarray}
It is easy to see that in the new basis, the problem of fluctuations between
0 and 1 particles has the same form as the interference one studied before
with level splitting $\delta =2t$ \cite{desint11,desint19,benz}.

Regarding the two-particle sector, the four relevant eigenstates for 
$U-U^{\prime }\rightarrow +\infty $ correspond to an even singlet and an odd
triplet. In the notation $|Sm\rangle $ of the total spin and its projection,
they are

\begin{eqnarray}
|00\rangle  &=&\frac{u}{\sqrt{2}}\left( d_{L\uparrow }^{\dagger
}d_{R\downarrow }^{\dagger }-d_{L\downarrow }^{\dagger }d_{R\uparrow
}^{\dagger }\right) |0\rangle\notag\\ 
	&&+\frac{v}{\sqrt{2}}\left( d_{L\uparrow
}^{\dagger }d_{L\downarrow }^{\dagger }+d_{R\uparrow }^{\dagger
}d_{R\downarrow }^{\dagger }\right) |0\rangle ,  \notag \\
|11\rangle  &=&d_{L\uparrow }^{\dagger }d_{R\uparrow }^{\dagger }|0\rangle ,\\
|10\rangle  &=&\frac{1}{\sqrt{2}}\left( d_{L\uparrow }^{\dagger
}d_{R\downarrow }^{\dagger }+d_{L\downarrow }^{\dagger }d_{R\uparrow
}^{\dagger }\right) |0\rangle ,  \notag \\
|1-1\rangle  &=&d_{L\downarrow }^{\dagger }d_{R\downarrow }^{\dagger
}|0\rangle .  \notag
	\label{states}
\end{eqnarray}
where to linear order in $t$, $u=1$, and $v=2\chi $ with $\chi
=t/(U-U^{\prime })$. The energy of the degenerate triplet is 
$E_{1}=2E_{d}+U^{\prime }$ and the corresponding one for the 
singlet is $E_{0}=E_{1}-J$ with $J=4t^{2}/(U-U^{\prime })$.

In the new basis, \ the problem for fluctuations between 1 and 2 particles
takes the  form  
\begin{eqnarray}
H_{1-2} &=&\sum_{\xi \sigma }E_{\xi }|\xi \sigma \rangle \langle \xi \sigma
|+\sum_{Sm_{S}}E_{S}|S,m\rangle \langle S,m|  \notag \\
&&+H_{c}+H_{V},  \label{hameff}
\end{eqnarray}
where  $\xi =\{e,o\} $ and $S=0,1$ and $-S\leqslant m\leqslant S$. In this
representation, the hybridization term of Eq. (\ref{hyb}) takes the form

\begin{eqnarray}
H_{V} &=&\sum_{k\nu \sigma }\sum_{\xi \sigma ^{\prime }}\sum_{Sm}V_{k\nu
}\,D_{\nu \sigma }^{Sm,\xi \sigma ^{\prime }}|S,m\rangle \langle \xi \sigma
^{\prime }|c_{\nu k\sigma }  \notag \\
&&+\text{H.c.},  \label{hv2}
\end{eqnarray}
where the matrix elements  $D_{\nu \sigma }^{Sm,\xi \sigma ^{\prime
}}=\langle S,m|d_{\nu \sigma }^{\dagger }|\xi \sigma ^{\prime }\rangle $,
are given in Appendix \ref{anca} together with a brief explanation of the
NCA treatment.

In what follows, we summarize the conventions and expressions of the charge and heat currents used in this work. Treating the system of Fig. \ref{scheme} as an interacting region coupled to conducting leads, the charge and energy currents are given by $J^{\nu}_{C}=-e\langle \dot{N}_{\nu} \rangle$,  $J^{\nu}_{E}=-\langle \dot{H}_{\nu} \rangle$ and $J^{\nu}_{Q}=J^{\nu}_{E}-\mu_{\nu}J^{\nu}_{C}$, with $e$ the absolute value of
the electronic charge and ${N}_{\nu}=\sum_{k \sigma }c^{\dagger }_{\nu k\sigma }c_{\nu k\sigma }$. Current conservation implies $J^{L}_{C,E}=-J^{R}_{C,E}$ and if $\mu_R=\mu_L$, $J^{L}_{Q}=-J^{R}_{Q}$.  In this paper we assume $\mu _{R}=\mu _{L}=0$. Therefore, the term
proportional to \ $\mu_{\nu}$ vanishes and we have $J^{\nu}_{Q}=J^{\nu}_{E}$. Furthermore, after a sign associated to each flow is chosen, the index $\nu$ can be dropped from the definition of the currents.
We take as positive the currents that flow from the left to the right lead.
Following a procedure similar to that described in the Appendix of Ref. 
\onlinecite{benz} the charge and heat currents are given by the following
expressions in terms of the physical Keldysh Green functions  obtained from
the NCA approximation

\begin{eqnarray}\label{mw-charge-heat-currents}
 J_{C}&=&\frac{ie}{h}\int d\omega~\mbox{Tr}\Big[ \big( \mathbf{\Gamma^{L}} f_{L}(\omega)-\mathbf{\Gamma^{R}} f_{R}(\omega) \big)\mathbf{G^{>}_{d}}(\omega)\notag \\  &&+\big( \mathbf{\Gamma^{L}} f_{L}(-\omega)-\mathbf{\Gamma^{R}} f_{R}(-\omega) \big)\mathbf{G^{<}_{d}}(\omega)\Big]\\
 J_{Q}&=&\frac{ie}{h}\int d\omega~\omega\mbox{Tr}\Big[ \big( \mathbf{\Gamma^{L}} f_{L}(\omega)-\mathbf{\Gamma^{R}} f_{R}(\omega) \big)\mathbf{G^{>}_{d}}(\omega)\notag \\  &&+\big( \mathbf{\Gamma^{L}} f_{L}(-\omega)-\mathbf{\Gamma^{R}} f_{R}(-\omega) \big)\mathbf{G^{<}_{d}}(\omega)\Big]
 \end{eqnarray}
where 
\begin{equation}
  \mathbf{\Gamma^{L}}=\Gamma_{L}\left(
    \begin{array}{cc}
      1&1\\
      1&1
    \end{array}
  \right),\quad
  \mathbf{\Gamma^{R}}=\Gamma_{R}\left(
    \begin{array}{cc}
      1&-1\\
      -1&1
    \end{array}
  \right).
\end{equation}
are the matrices that couple the even and odd levels to the left and right reservoirs. Furthermore $\Gamma_{\nu}=\pi V^{2}_{\nu}/D$ where $2D$ represents the conduction band width and $f_{\nu}(\omega)=[1+ \mbox{exp}[(\omega- \mu_{\nu})]/k_{B}T_{\nu} ]^{-1}$ is the Fermi function. 
We note that since $\mathbf{\Gamma^{L}}$
and $\mathbf{\Gamma^{R}}$ are not proportional, it is not possible to eliminate the nonequilibrium Green functions in the expressions of the current, and even in the linear response regime, we have to use the full non-equilibrium formalism
and obtain electric and thermal conductances deriving
numerically the corresponding currents. In addition, 
NRG approaches in the linear response regime, in which the 
conductances are related to the spectral density and its derivative \cite{costi}, are not applicable in this case.

In Ref. [\onlinecite{benz}] it was proved that the NCA is a charge current conserving approximation for a two level Anderson model. Here we have verified that the same is true for the energy (and heat in the present case) current.

For $t=0$, the system is disconnected
and it is not possible to transfer particles between both dots.
Therefore, the electric current vanishes (but not necessarily 
the heat current as it is shown in Fig. \ref{jq2}).
The use of the above expression for the electric current becomes numerically inaccurate for small $t$, because 
it requires the cancellation of large positive and negative terms. 
This fact has been also found in the
interference problem \cite{desint11,desint19}. To avoid this problem, we have
derived an alternative expression for the charge current which is explicitly proportional to $t$ and therefore, is more accurate for small $t$.
The derivation is included in Appendix \ref{current-delta}. The result is

\begin{eqnarray}
J_{C}&=&\frac{2\pi et}{h}\sum_{\sigma}\int {d\omega }~\mbox{Re}\Big(G_{eo,\sigma}^{<}(\omega)\Big)\notag\\
&=&\frac{2\pi et}{h}\sum_{\sigma}\int {d\omega }~\mbox{Re}\Big(G_{LR,\sigma}^{<}(\omega)\Big).  \label{currd}
\end{eqnarray}

\section{Results}

\label{res}

\subsection{Fluctuations between 0 and 1 particles in the system}

\label{res01}

Here we assume that not only $U$ is very large but also 
$U^\prime$ is sufficiently large to avoid occupancy of more than
one particle in the double-dot system. The case in which the 
total number of particles in the system fluctuates between 3 and 4 particles can be mapped to the specific case treated in this section by an electron-hole transformation and change of sign of 
all the operators at the left or right part of the system.

We take the Fermi energy $\epsilon_F=0$ as the origin of one-particle energies and assume $E_d <0$.
The problem becomes equivalent to that of transport through 
one dot or a molecule with two levels which interfere
destructively in the transport \cite{desint19,desint11,benz}.
Several mappings were discussed in Ref. \onlinecite{desint19}.
In the limit $t \rightarrow 0$ the spectral density for each dot
tends to that of the SU(4) Anderson model (with spin and 
``orbital'' degeneracy), with a narrow peak of half width
$\sim T_K^\text{SU(4)}$ at energy also $\sim T_K^\text{SU(4)}$ above the Fermi level 
and a broad charge transfer peak centered at $E_d$ and width \cite{anchos} $4 \Delta$, where $T_K^\text{SU(4)}$ is the Kondo temperature
of the SU(4) Anderson model, and $\Delta$ is the resonant level width.
The spectral density of the SU(4) Anderson model 
showing both peaks is represented 
for example in Fig. 1 of Ref. \onlinecite{restor}.
As $t$ increases, the even and odd one-particle states split
being the former the one of lowest energy 
(see Sec. \ref{model}).
For splitting $\delta > T_K^\text{SU(4)}$ the density of states of the even state tends to
the corresponding one of the SU(2) model with a charge transfer peak of width $2 \Delta$ \cite{anchos} and a Kondo peak 
slightly above $\epsilon_F=0$ of half width 
$T_K^\text{SU(2)} \ll T_K^\text{SU(4)}$ \cite{su42,restor}.  The spectral density of the odd state displays a peak at energy $2t$ instead of the Kondo peak \cite{su42}.

The electrical conductance $G=dI/dV$ of the model has been studied before
\cite{desint11}. It vanishes (as expected) for $t=0$ and for 
large $t$ and zero temperature, 
it tends to the value $G_0=2 e^2/h$ characteristic
of the SU(2) Kondo model, with a crossover when 
$\delta \sim T_K^\text{SU(4)}$. The decay of $G$ with
increasing temperature resembles that expected for an
SU(4) or an SU(2) model depending on the ratio
$\delta_4 = \delta / T_K^\text{SU(4)}$ \cite{su42}.
Near the crossover also the occupancies of the even and odd combinations change from being
nearly equal (slightly below 1/2 adding both spins) in the SU(4) regime, to be dominated by the even one in the SU(2) regime \cite{desint11}.

To study thermoelectric effects, we take 
$\Gamma = \Gamma_L + \Gamma_R=1$ as the unit of energy,
$E_e=E_d=-4$ and $E_o=E_e+\delta$, where $\delta=2t$. 
We also choose the half band width $D=10$. 
For these parameters, from the equation $G(T_K)/G(0)=1/2$,
for $\delta \rightarrow 0$, 
we estimate $T_K^\text{SU(4)}=0.011$, 
using the variational equation \cite{su42,desint11}

\begin{equation}
T_{K}=\left\{ (D+\delta )D\exp \left[ \pi E_{d}/(2\Delta )\right]
+\delta ^{2}/4\right\} ^{1/2}-\delta /2,  
\label{tkd}
\end{equation}
 with $\delta=0$ and this value is used in the definition of 
 $\delta_4 = \delta / T_K^\text{SU(4)}$ below.
 The Kondo temperature that enters in Figs. 2 to 7 is given by Eq. (\ref{tkd}).

\begin{figure}[h!]
\centering
\includegraphics[width=0.45\textwidth]{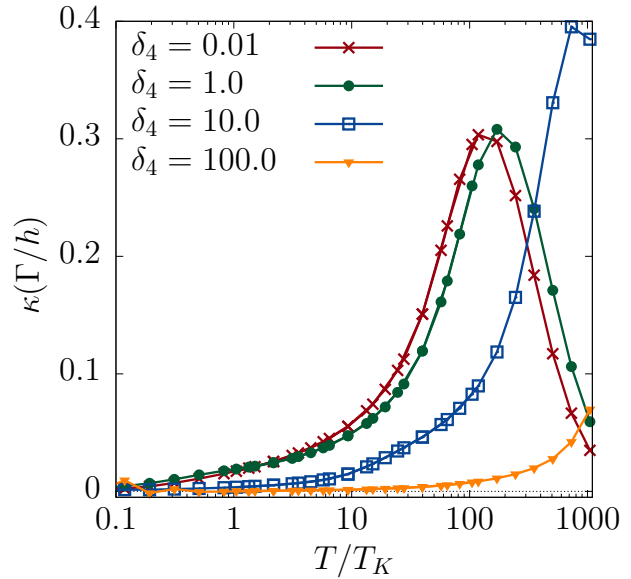}  
\caption{(Color online) Thermal conductance as a function
of temperature for several values of $\delta_4$.}
\label{k}
\end{figure}

In Fig. \ref{k} we show the thermal conductance $\kappa$ 
in the linear response regime as a function of temperature 
for several splittings. In contrast to the electrical conductance, the thermal conductance increases strongly with
increasing temperature for temperatures of the order of 
$T_K^\text{SU(4)}$ in particular for small $\delta_4$ (in the SU(4) regime, where the Kondo peak is clearly above
the Fermi energy). As $\delta_4$ increases beyond 1, the system enters the SU(2) regime, where the spectral density 
near the Fermi energy is more electron-hole symmetric, and
$\kappa$ is markedly reduced.
 
For temperatures above the charge-transfer gap $|E_d|$,
the thermal conductance is expected to decrease. 
This seems to happen at energies below 
$|E_d| \sim 364 T_K^\text{SU(4)}$ for
small $\delta_4$ (for which $T_K \sim T_K^\text{SU(4)}$) . This is probably an effect of destructive interference.

\begin{figure}[h!]
\centering
\includegraphics[width=0.45\textwidth]{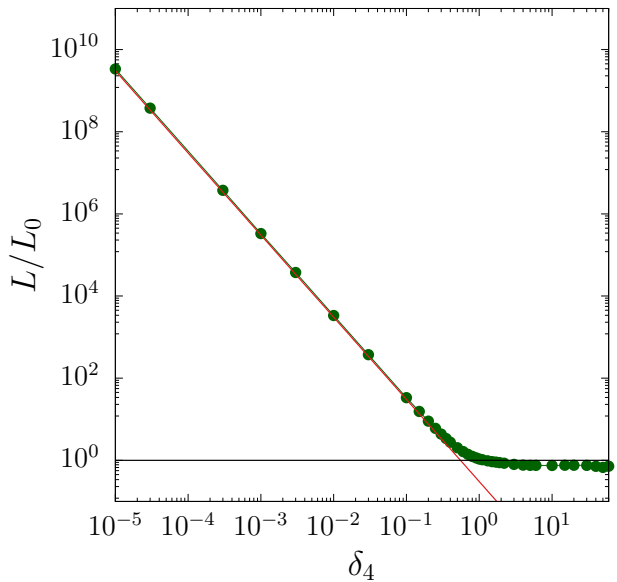}  
\caption{(Color online) Lorentz number as a function of  
$\delta_4$, at $T=0.8 T_K$. }
\label{l}
\end{figure}

A remarkable result is that while the Lorenz number 
tends  approximately to the free electron value 
$L_0=\pi^2/3 (k_b/e)^2 $, in the SU(2) regime, it diverges as 
$ \delta_4^{-2}$
as $\delta_4 \rightarrow 0$. 
This fact is illustrated in Fig. \ref{l}.
We are not aware of other systems with a similar property.
This is a peculiar violation of the  Wiedemann-Franz  law. Previous violations obtained by 
calculations in nanoscopic systems
were by factors below 5 \cite{Kubala,burkle}.

\begin{figure}[h!]
\centering
\includegraphics[width=0.45\textwidth]{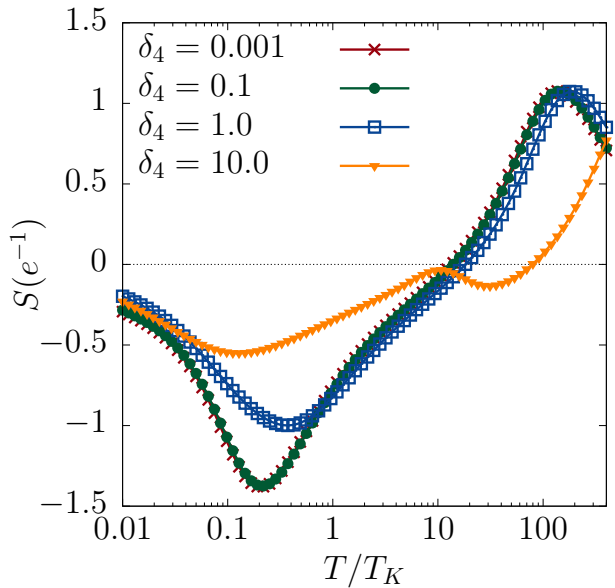}  
\caption{(Color online) Thermopower as a function
of temperature for several values of $\delta_4$.
}
\label{s}
\end{figure}

In Fig. \ref{s} we show the behavior of the Seebeck
coefficient $S$, for several values of $\delta_4$.
In the SU(4) regime of small $\delta_4$, $S$ behaves similarly as the corresponding SU(4) result without
destructive interference \cite{see}. At low temperatures,
the thermopower is dominated by the Kondo peak in the spectral density al low energies which lies above the Fermi energy, thus, it is negative, characteristic of hole transport.
Instead at temperatures of the order of the charge-transfer energy $|E_d|$, $S$ is dominated by the charge transfer peak, which is centered at $E_d$ and is below the Fermi energy.
Therefore, $S$ changes sign at intermediate temperatures.
However, while the maximum value of $|S|$ in the 
standard SU(4) case is near 1 $k_B/\Gamma$,
in our case with destructive interference it is about 
40\% larger. Since in this case, both the electrical and the thermal current vanish in the limit $\delta_4 \rightarrow 0$,
the origin of this difference is unclear.

Increasing $\delta_4$, the low-energy minimum
approaches zero because the Kondo peak moves towards 
the Fermi energy. One also expects a negative contribution 
at energies of the order of $\delta$ due to the peak
at energy $\delta$ in the spectral density of the odd state, 
which competes with the positive contribution of the charge transfer peaks at energies $E_d$ and $E_d+\delta$ below
the Fermi energy.

\begin{figure}[h!]
\centering
\includegraphics[width=0.45\textwidth]{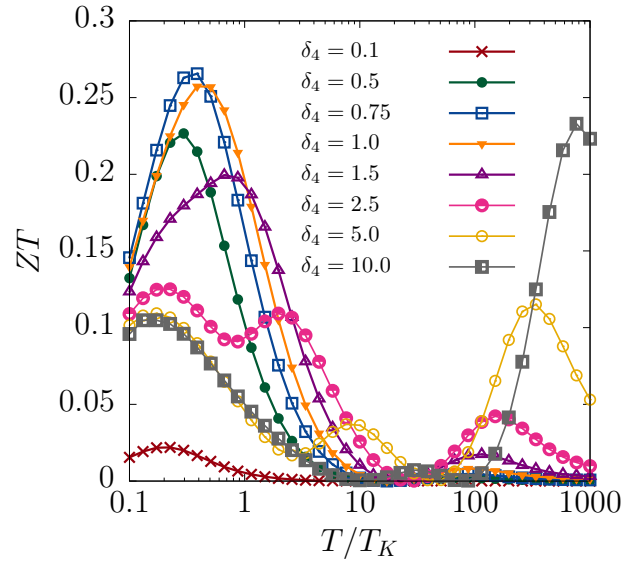}  
\caption{(Color online) Figure of merit as a function of temperature.
}
\label{zt}
\end{figure}

In Fig. \ref{zt} we show the figure of merit,
which can be relevant for possible applications
\cite{bene,balan}.
One can see that the most convenient choice to increase
$ZT$ is the intermediate regime $\delta_4 \sim 1$, where 
$ZT$ can reach 1/4.

\begin{figure}[h!]
\centering
\includegraphics[width=0.45\textwidth]{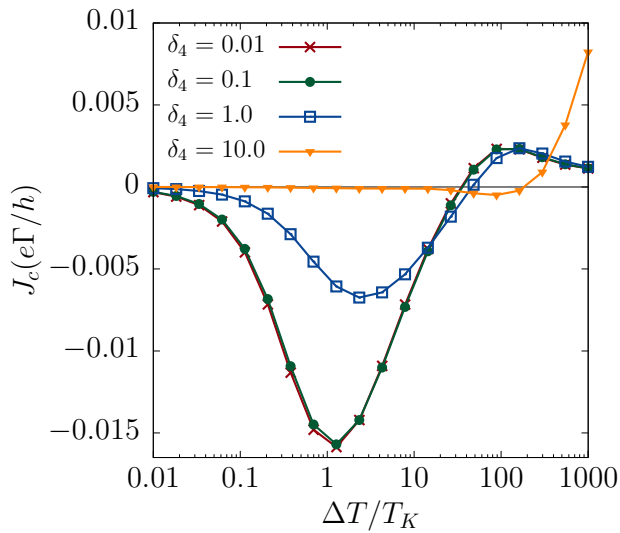}  
\caption{(Color online) Electrical current as a function of the temperature difference $\Delta T=T_L-T_R$,
keeping $T_R=0.1 T_K$. For clarity, curves are rescaled with factors: $10^6$, $10^2$, and $10$ for $\delta_4$ =0.01, 0.1 and 10, respectively.}
\label{jc}
\end{figure}

In Fig. \ref{jc} we show the thermoelectric current 
induced by a difference in the temperature between
the left and right leads $\Delta T=T_L-T_R$. In spite of 
the obviously different physical situation, the curve has 
some features that resemble the thermopower shown above.
For small $\delta_4$ and $\Delta T$, negatively 
charged electrons are excited
in the Kondo peak above the Fermi level at the left dot 
and move to the right, originating a positive particle current and a negative charge current. When the 
left lead reaches temperatures of the order of the charge transfer peak $|E_d|$, holes are promoted in this lead which can be filled from electrons of the right lead, 
leading to negative particle current and positive 
charge current. 

Similar arguments can be followed for large $\delta_4$
[in the SU(2) regime], following a reasoning analogous 
to that made above for the thermopower. However, in this case, the effect of increasing $\delta_4$ seem more dramatic.

\begin{figure}[h!]
\centering
\includegraphics[width=0.45\textwidth]{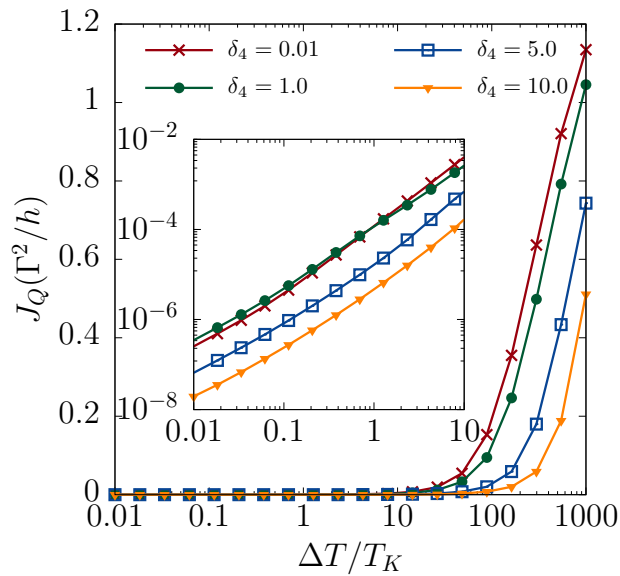}  
	\caption{(Color online) Heat current as a function of the temperature difference $\Delta T=T_L-T_R$, keeping $T_R=0.1 T_K$. (Inset) Zoom of the region with $\Delta T/T_K$ lower than $10$.}
\label{jq}
\end{figure}

In Fig. \ref{jq} we represent the corresponding heat 
current for the same situation as the previous figure. 
It seems to reach important values only at energies near
the charge-transfer energy $|E_d|$ and particularly 
for small $\delta_4$.

\subsection{Fluctuations between 1 and 2 particles in the system}

\label{res12}

The problem for fluctuations between 2 and 3 particles can be mapped into the corresponding one for 1 and 2 particles 
using the electron-hole transformation mentioned at the beginning of Sec. \ref{res01}. In the following we refer to the latter case.

The configuration with two particles consists of a singlet even under inversion and an odd triplet separated by an energy difference which is small for large $U$, $E_0=E_1-J$ with $J=4t^2/(U-U^{\prime})$.
The configuration with one particle consists of an 
even doublet and 
an odd one. The two-particle configuration is that of lowest energy for small $t$ and $U^\prime$.

A model in which a singlet and a triplet are mixed with 
a doublet, all with the same parity has been studied first in the context of Tm impurities in a cubic crystal field \cite{allub}
and more recently to explain experiments \cite{roch} in a 
single-molecule quantum dot \cite{sitri1,sitri2,serge},
This model has a quantum phase transition (QPT) which separates 
two regions with a doublet or a singlet ground
state. This transition is accurately described by the 
NRG \cite{allub}.
With the NCA a change in the spectral density near the 
Fermi level, accompanied by a change in the occupancy of
the triplet and singlet auxiliary bosons is found \cite{sitri2}.

In the case of two dots interacting via a Heisenberg interaction, there is also a QPT which turns to a crossover 
when hopping between the dots is included \cite{zara06,dong02}. It is likely that this QPT is 
related to the previous one.

In our model we find that for $\chi=0$ (degeneracy of singlet  
and triplet states of the two-particle localized configuration)
the triplet boson dominates, but small values of $\chi$ 
(of the order of 0.001) induce a crossover to  the dominance 
of the singlet boson at low temperatures and the disappearance
of the Kondo peak at the Fermi level.

\begin{figure}[h!]
\centering
\includegraphics[width=0.5\textwidth]{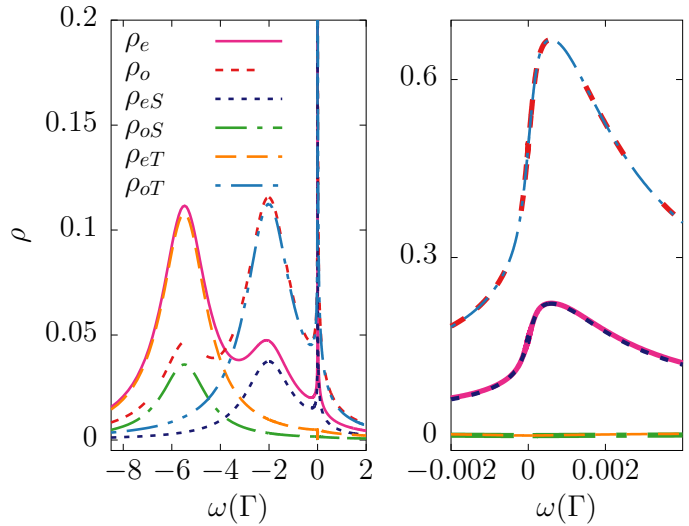}  
	\caption{(Color online) Contribution of the two-particle singlet and triplet states to the spectral density of the even and odd states for $t=1.5$, $U^\prime=\chi=0$ and $T=10^{-4}$. Right panel: zoom of the region of the Kondo peak.} 
\label{rho}
\end{figure}

As a basis for our study we take $\Gamma_L=\Gamma_R=1$ as 
the unit of energy and $E_d=-3.5$.
We also take for the moment $\chi=0$.
As before we choose the origin of one-particle energies 
at the Fermi level $\epsilon_F=0$.

With the above assumptions, the energies for the 
even and odd one-particle states are $E_e=E_d-t$, $E_o=E_d+t$.
The spectral density of these states is shown in Fig. \ref{rho}
for a typical case in the regime of small $t$ and $U^\prime$.
The Kondo peak lies slightly above the Fermi energy and the charge transfer peak is split by $t$. The even (odd) state has 
larger weight at $E_d-t$ ($E_d+t$). 
This is explained as follows. The ground state is composed mainly
of the odd triplet with an even and an odd particle and 
total energy near $2 E_d$. Destroying the even particle 
leaves the odd one with energy  $E_d+t$. Therefore, the energy
difference is $E_d-t$. Interchanging even and odd states 
the reasoning is similar with a change in the sign in 
front of $t$.
Due to the even parity of the two-particle singlet, 
the contribution of this singlet to the split peaks
is inverted with respect to the contribution of the
triplets, which is the dominant one.
Except for the splitting,
the spectral density is qualitative similar to that expected for the ordinary Anderson model at large $U$. 

\begin{figure}[h!]
\centering
\includegraphics[width=0.45\textwidth]{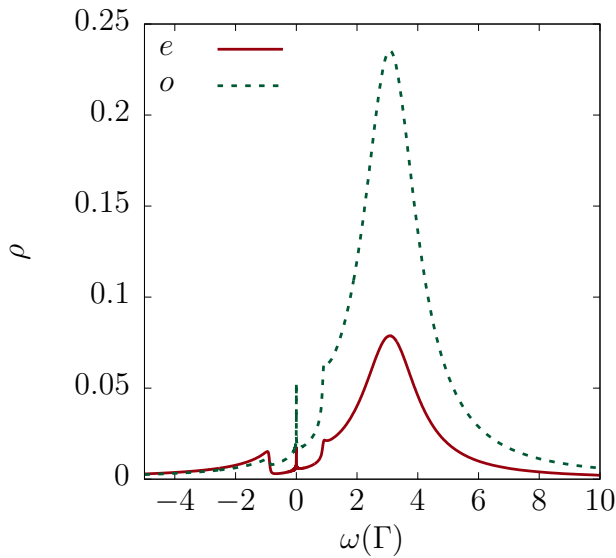}  
\caption{(Color online) Spectral density of the even and odd  states for $t=0.5$, $U^\prime=6$  and $T=10^{-4}$.
}
\label{rhoup}
\end{figure}

When $t$ reaches $|E_d|$, there is a change of valence 
and the one-particle even state becomes the ground state.
The same happens with increasing $U^\prime$. 
An example for the spectral density for large $U^\prime$
is shown in Fig. \ref{rhoup}. It has three main features:
the charge transfer peak, now at positive energies with respect
to the Fermi level, approximately at $E_d+U^\prime+t$ 
(which is the difference in energy between the two-particle 
states at $2E_d+U^\prime$ and the ground one-particle 
state at $E_d-t$), a narrow Kondo peak at the Fermi energy
(very slightly displaced to negative energies) and two inelastic features at $\pm 2t$ related to higher order processes
involving transitions between even and odd one-particle states.

\begin{figure}[h!]
\centering
\includegraphics[width=0.45\textwidth]{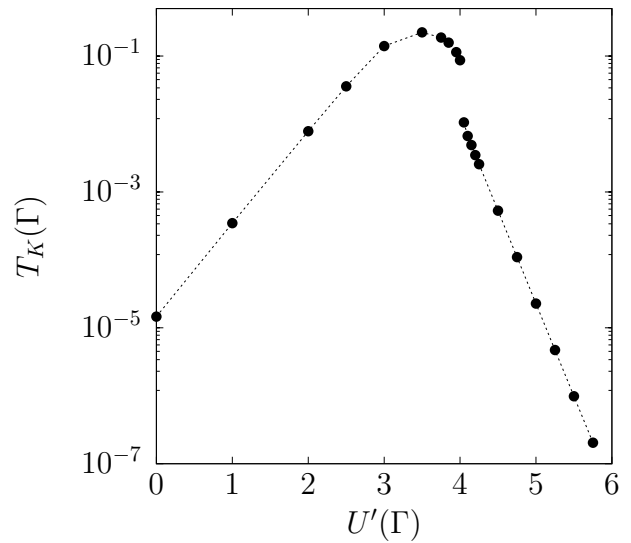}  
\caption{(Color online) 
Energy scale $T_K$ obtained from the pseudoconductance (see text)
as a function of $U^\prime$ for $t=0$
}
\label{es}
\end{figure}

To study the behavior of different quantities, it is useful 
to define the energy scale related to the half width at half maximum of the Kondo peak, which is one way to define the 
Kondo temperature $T_K$. However, we had found that
the result of fitting this resonance at the Fermi level depends on the range of the fit and it can vary within a factor 2 depending on this range \cite{width}. Therefore, 
to determine $T_K$ we have used the 
``pseudoconductance'' $G_p(T)$,
which is the conductance through one quantum dot with
the total spectral density of our system, obtaining 
$T_K$ from the expression $G_p(T_K)=G_p(0)/2$.
In Fig. \ref{es} we represent $T_K$ obtained in this way
as a function of $U^\prime$ in a logarithmic scale.
In agreement with previous calculations using NRG \cite{mit06},
the energy scale increases strongly at the intermediate 
valence regime and decreases in going to the integer valence limits of two particles in the system for small $U^\prime$ 
or 1 particle for large $U^\prime$.

\begin{figure}[h!]
\centering
\includegraphics[width=0.45\textwidth]{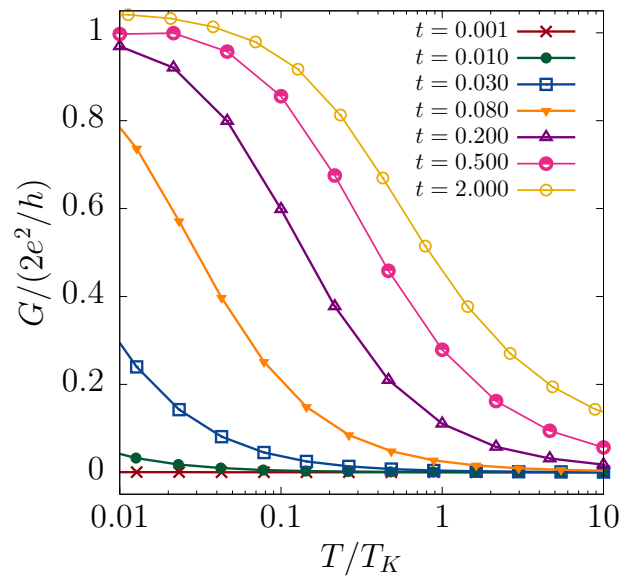}  
\caption{(Color online) Electrical conductance as a function of temperature for $U^\prime=0$  and several values of $t$.
}
\label{g}
\end{figure}

The temperature dependence of the real conductance $G(T)$ 
through
the system is shown in Fig. \ref{g} in the regime of 
total occupancy near 2 in the system. For $t \gtrsim 0.5$,
$G(T_K) \sim G(0)/2$ indicating that the conductance
is dominated by the total density of states, as expected
in general. However, for low $t$ the behavior is dominated by
the destructive interference. Not only $G(T) \rightarrow 0$ 
for $t \rightarrow 0$, but also a new small energy scale 
$T^*$ appears
which determines the temperature at which $G(T)$ begins 
to fall with increasing $T$. For $t \sim 10^{-2}$, 
$T^*\sim 10^{-2} T_K$.

\begin{figure}[h!]
\centering
\includegraphics[width=0.45\textwidth]{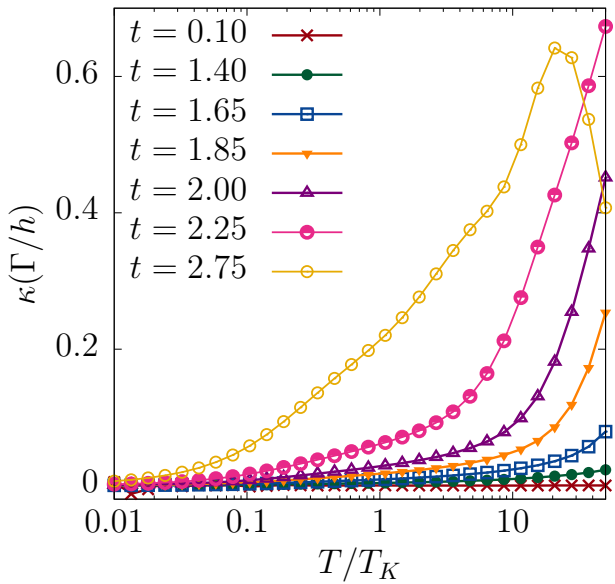}  
\caption{(Color online) Thermal conductance as a function of 
temperature for $U^\prime=0$  and several values of $t$.
}
\label{k2}
\end{figure}

In Fig. \ref{k2} we represent the thermal 
conductance $\kappa$
as a function of temperature for the same parameters
as in Fig. \ref{g}. It increases monotonically with 
$T$ until temperatures greater than the charge-transfer energy are overcome. It is also clear that 
$\kappa$ increases substantially as the intermediate-valence
regime between total occupancies 2 and 1 is approached.

\begin{figure}[h!]
\centering
\includegraphics[width=0.45\textwidth]{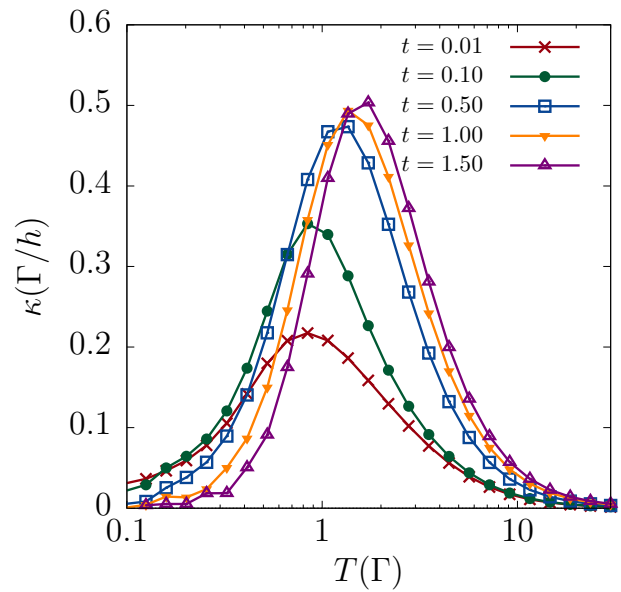}  
	\caption{(Color online) Thermal conductance as a function of 
	temperature for $U^\prime=6$  and several values of $t$. 
}
\label{k2p}
\end{figure}

The thermal conductance for relatively large $U^\prime=6$, 
for which the occupancy of the system is slightly above 1, 
is shown in Fig. \ref{k2p}. There is a maximum at 
temperatures of the order of the charge-transfer energy. 
As expected, $\kappa$ increases with increasing $t$,
but the effect is not linear in $t$. 
Note that in contrast to the previous figures, there is a significant thermal conductance at $t=0$ due to the effect 
of $U^\prime$. This fact is known from previous calculations
in a spinless model \cite{san1,san2,ruoko,yada,heat}.

\begin{figure}[h!]
\centering
\includegraphics[width=0.45\textwidth]{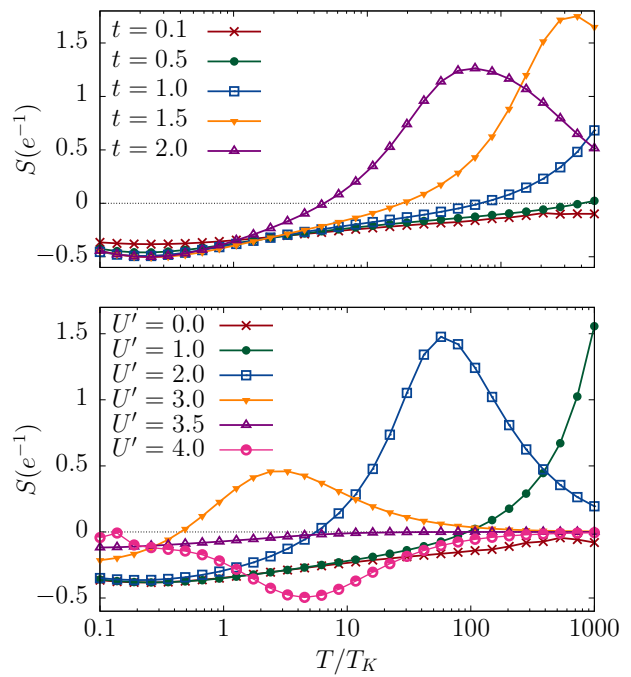}  
	\caption{(Color online) Thermopower as a function of temperature. Top panel: for $U^\prime=0$ and several values of $t$. Bottom panel: for $t=0.1$ and several values of $U^{\prime}$. 
}
\label{s2}
\end{figure}

In Fig. \ref{s2} top panel, we show the thermopower as a function of temperature in the regime of total occupancy near 2.
Since the spectral density is qualitatively similar to that
of Sec. \ref{res01} in the SU(4) regime, 
the figure has same features in common with Fig. \ref{s}.
In particular, there is a dip at low temperatures due to the 
Kondo peak slightly above the Fermi level and a peak at 
the charge-transfer energy $|E_d|-t$ due to the charge-transfer peak in the spectral density next to the Fermi level. The negative interference for $t \rightarrow 0$ 
tends to suppress both, the thermal and electrical conductances but does not affect the thermopower in a 
marked way. 

In Fig. \ref{s2} bottom panel, we show the thermopower as a function of temperature for several values of $U^\prime$. The 
peak at the temperature corresponding to the 
charge-transfer energy $\sim |E_d|-U^\prime$ displaces to lower temperature with increasing $U^\prime$. 
Entering the regime of total occupancy 1, for large 
$U^\prime$, the thermopower
is strongly reduced, particularly at small temperatures, 
because the Kondo peak is nearly symmetric and with small
intensity (see Fig. \ref{rhoup}). In addition since
now the charge-transfer peak in the spectral density lies at positive energies, the corresponding feature in the thermopower becomes negative.

\begin{figure}[h!]
\centering
\includegraphics[width=0.45\textwidth]{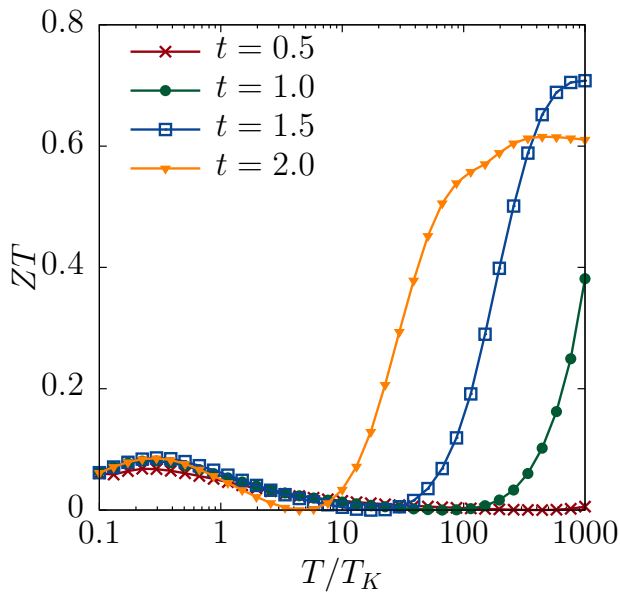}  
\caption{(Color online) Figure of merit as a function of temperature for $U^\prime=0$ and several values of $t$.
}
\label{zt_o12}
\end{figure}

The figure of merit $ZT$  reaches values near 0.7
for temperatures of the order of the charge gap and $t \sim 1.5$, see Fig. \ref{zt_o12}.

%\begin{figure}[h!]
%\centering
%\includegraphics[width=0.45\textwidth]{S_o12_Up.eps}  
%\caption{(Color online) Thermopower as a function
%of temperature for $t=0.1$ and several  values of 
%$U^\prime$.
%}
%\label{s2up}
%\end{figure}

\begin{figure}[h!]
\centering
\includegraphics[width=0.45\textwidth]{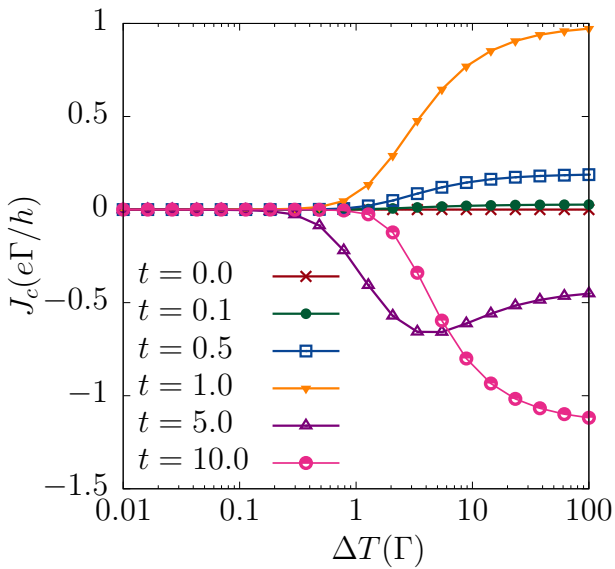}  
\caption{(Color online) Electrical current as a function of the temperature difference $\Delta T=T_L-T_R$,
keeping $T_R=10^{-3}$ for $U^\prime=0$ and several values 
of $t$.}
\label{jc2}
\end{figure}

The thermoelectric current, shown in Fig. \ref{jc2} has 
qualitative features similar to the thermopower shown in Fig. \ref{s2} for small $t$. In particular, since there is a
charge transfer peak below the Fermi level, at temperatures at which holes are excited in the left lead, these holes
can move to the right lead leading to a positive charge current. However for $t>|E_d|$ there is a valence 
crossover from occupancy near 2 to 1 in the system, 
and the charge transfer peak moves above the Fermi level, leading to a change in the sign of the electrical current.
Increasing $U^\prime$ has a similar effect (not shown).

\begin{figure}[h!]
\centering
\includegraphics[width=0.45\textwidth]{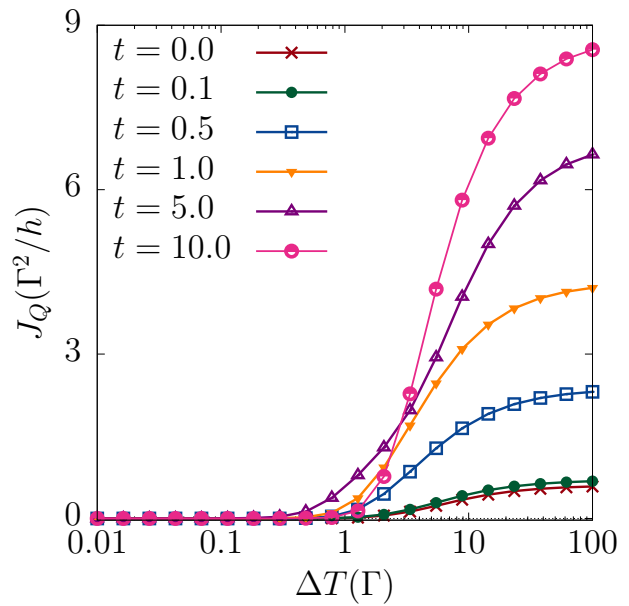}  
\caption{(Color online) Heat current as a function of the temperature difference $\Delta T=T_L-T_R$,
	keeping $T_R=10^{-3}$ for $U^\prime=0$ and several values of $t$.
}
\label{jq2}
\end{figure}

\begin{figure}[h!]
\centering
\includegraphics[width=0.45\textwidth,height=9cm]{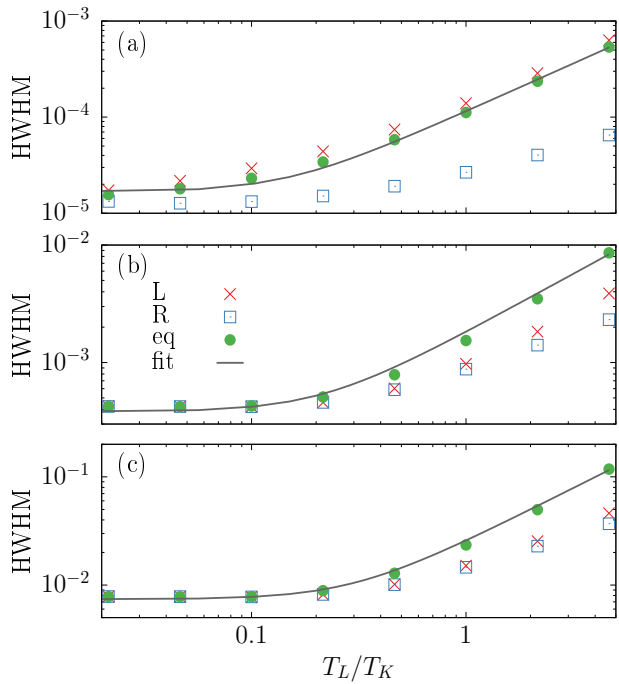}  
\caption{(Color online) half widths at half maximum 
of the spectral densities of the left (crosses) and right 
(squares) dots 
as a function of $T_L$, keeping $T_R=0.01T_K$ for $U^\prime=0$ and (a) $t=0.0$, (b) $t=1.0$ and (c) $t=1.85$.
The green dots correspond to both leads at temperature $T_L$.
}
\label{widths}
\end{figure}

In Fig. \ref{jq2} we display the heat current for similar 
parameters as the previous figure. A noticeable effect
is that even for $t=0$, for a difference of temperature
$\Delta T$ higher than $\Gamma$, 
there is a small but significant heat current 
due to the effect of correlations. Previous works
have demonstrated this effect as a consequence 
of $U^\prime$ in a spinless model \cite{san1,san2,ruoko,yada,heat}.
Here we obtain that a similar effect takes place also as a consequence of large on-site repulsion $U$, although
our results for the thermal conductance (Fig. \ref{k2})
give negligible values for $t=U^\prime=0$.
We return to this point later, when the width of the spectral density of both dots is discussed (Fig. \ref{widths}). 
The heat current increases as the occupancy is reduced 
from 2 to 1 at high temperatures.

When $U^\prime$ is included, as expected 
from previous work \cite{heat}, the heat transport at $t=0$ is larger in the intermediate 
valence regime, reaching values near $J_Q=0.7 (\Gamma^2/h)$  for $U^\prime=4$ and fixed $\Delta T=10$. For $U^\prime=0$,  
$J_Q$ is slightly below $0.2 (\Gamma^2/h)$
For the sake of brevity we do not include here our
studies for finite $U^\prime$. 
As in previous studies \cite{san1,san2,ruoko,yada,heat}, the heat current is large for $U^\prime \neq 0$ even at $t=0$.
Instead, slave boson approaches give always $J_Q=0$ for $t=0$.

In contrast to Sierra \textit{et al.} \cite{sierra},
we do not find a regime of decreasing $J_Q$ with
increasing $\Delta T$. A discussion of the possible reasons 
for this discrepancy is left to Sec. \ref{sum}.

As in the case of fluctuations between 0 and 1 particles
presented in Sec. \ref{res01}, we also find strong violations of the  Wiedemann-Franz  law. The ratio 
$J_Q/(T J_c)$ ranges from $\sim 0.1/e^2$ for $\Delta T=100$
to $\rightarrow \infty$ for parameters for which 
$J_c \rightarrow 0$ (not shown).

In Fig. \ref{widths} we show the evolution of the widths 
of the spectral densities at both dots as a function of the 
temperature of the left (hot) lead $T_L$ keeping the temperature
of the right (cold) lead $T_R$ constant. For the physical discussion below,  we also show the result of the equilibrium situation,
when the temperature of both leads is increased simultaneously
($T_R=T_L$ instead of fixed $T_R$).
In this case, the half width at half maximum (HWHM) of the spectral density is expected to vary with temperature 
according to 

\begin{equation}
\text{HWHM} = \sqrt{(\alpha T)^2 + T_{K}^{2}},
\label{wi}
\end{equation}
where $\alpha$ is fitting parameter \cite{vibra,Otte08}. 
As it happened previously \cite{vibra}, the NCA results
are well fitted by this expression 
(full lines in Fig. \ref{widths}), with values of $\alpha$
in the range between 3 and 8, similar to values previously 
obtained \cite{vibra,Otte08}.

For $t=0$ both dots are disconnected by the one-particle terms 
of the Hamiltonian. One would expect then that each dot reaches the temperature of the corresponding lead in the absence of interatomic interactions 
($U^\prime=0$). The HWHM of the left dot is slightly 
above the equilibrium value at temperature $T_L$. Except for 
this small deviation, the result agrees with the expectations.
Instead, one would expect that the HWHM of the right dot
remains at the corresponding value for $T_R$. However, it increases 
moderately with increasing $T_L$. 
One knows that in presence of interatomic
interactions, in particular $U^\prime$, 
there is heat transfer 
between the dots \cite{san1,san2,ruoko,yada,heat}, 
and this is consistent with the different results
presented here. 
However, it is not clear that a similar effect can exist 
for $U^\prime=0$. In might be an effect of the entanglement
between the dots introduced by the change of basis and neglect of excited states combined with the NCA approach.

As $t$ increases, one expects an interchange of heat between
the dots and that the HWHMs are intermediate between those
corresponding to the equilibrium ones, replacing $T$ by 
either $T_L$ or $T_R$ in Eq. (\ref{wi}). This is indeed what 
happens. For large hopping, the results of Fig. \ref{widths}
indicate that both dots tend to be at the same temperature,
intermediate to those of both leads.

\subsubsection{Effect of triplet-singlet splitting}

\label{chi}

In this section, we examine the effect of a finite triplet-singlet
splitting on the thermoelectric properties out of equilibrium.

\begin{figure}[h!]
\centering
\includegraphics[width=0.5\textwidth]{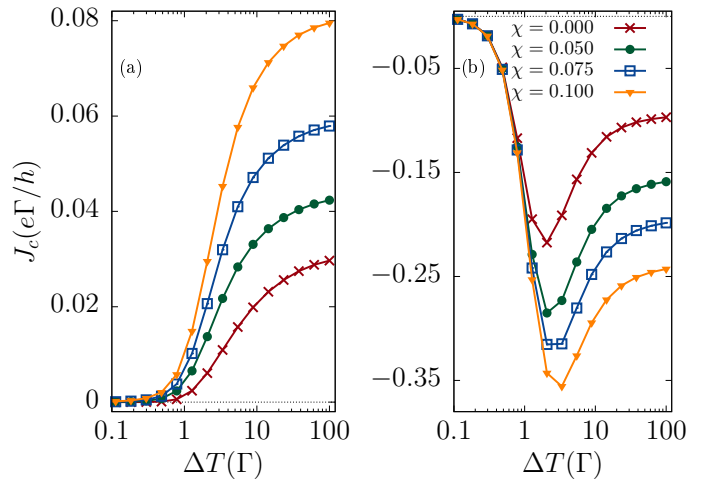}  
\caption{(Color online) Electrical current as a function of the temperature difference $\Delta T=T_L-T_R$,
	keeping $T_R=0.1T_K$ for  $t=0.1$, different  values of 
	$\chi$ and (a) $U^\prime=0$ and (b) $U^\prime=6$.
}
\label{jcchi}
\end{figure}

The effect of a finite triplet-singlet
splitting on the electrical current produced by a finite temperature
difference between the leads is shown in Fig. \ref{jcchi} (a).  
The thermoelectric current is enhanced at moderate temperatures.
The same happens at large $U^\prime$. However, in this case
since the charge-transfer peak is above the Fermi energy, electrons
flow from the cold to the hot lead, leading to a negative
particle current and a positive charge current.
The heat current is also enhanced.

\section{Summary and discussion}

\label{sum}

In summary, we have studied an inversion symmetric 
DQD system in series, with each dot connected to its own lead, under a thermal bias.
We have used the non-crossing approximation in the limit of
very large on-site Coulomb repulsion. In this limit,
the problem can be divided into fluctuations between 
0 and 1 electron (regime I) in the DQD, 
or between 1 and 2 electrons (regime 2).
Other fillings are related to the previous ones by electron-hole symmetry.

The behavior of the different quantities studied can be understood from the energy dependence of the total spectral density $\rho(\omega)$ at equilibrium and low temperatures. While the charge conductance decreases in general with increasing temperature, 
the thermal conductance is maximum at temperatures of the order of the charge-transfer energy (the difference between the energies of the configurations with integer number 
of particles in the DQD). For interdot hopping $t=0$,
the heat conductance is negligible if the interdot repulsion
$U^\prime$ is also zero, but increases significantly
for finite $U^\prime$. This fact is known from previous 
calculations in simplified systems\cite{san1,san2,ruoko,yada,heat}, but is 
not reproduced by the 
simplest SBMFA,
because in the latter, the low-energy physics is reduced to
an effective non-interacting system and the effect of 
$U^\prime$ is not included.

In contrast to previous work \cite{sierra}, we do not find
regions of negative differential thermal conductance.
The reason of the discrepancy is the following. 
As explained in Ref. \onlinecite{sierra}, within the 
SBMFA the effective Kondo temperature of the hot lead is
decreased with increasing temperature of the lead, 
and for very large temperatures the transport through
the double dot is blocked. Therefore, the thermal transport is inhibited as the temperature of the hot lead is increased. We believe that this an
artifact of the SBMFA, of similar character 
as the artificial decoupling of the impurity 
with increasing temperature \cite{hewson} or magnetic field \cite{lady} found previously.
Instead, the NCA actually improves with increasing 
temperature and the width of the spectral density
follows the expected temperature dependence given by 
Eq. (\ref{wi}) (see Fig. \ref{widths} and Ref. \onlinecite{vibra}).

The sign of the charge current and the thermopower can be inferred from the positions and magnitude of the different 
peaks in $\rho(\omega)$, one related to the Kondo effect, near the Fermi energy and one or two related with charge transfer-processes, which can be above or below the Fermi energy.
We find sign reversals of the thermoelectric current, in agreement with previous work \cite{sierra}.
In regime I, for small interdot hopping $t$, the thermopower $S$ 
reaches a maximum absolute value at temperatures $T$ below 
the Kondo energy scale $T_K$ near $1.5/e$ and the Lorentz ratio diverges for $t \rightarrow 0$, indicating a strong 
violation of the Wiedemann-Franz law, unexpected 
from measurements in a large number of systems \cite{kumar93}.
Previous deviations
of this law calculated in mesoscopic systems were much more 
moderate \cite{Kubala,burkle}. 
There is also a maximum of $S$ near 1 
at high temperatures. This value is much higher than
the corresponding one for a single-level quantum dot \cite{costi,see}.  
The figure of merit $ZT$ reaches values of the order of 
0.25 at $T \sim T_K$ for moderate $t$ and at temperatures 
of the order of $2t$ for high $t$.

In regime II the thermopower reaches also high 
values but at temperatures of the order of the charge-transfer energy. The Lorentz ratio can be also very large or very small depending on the values of the parameters. On the other hand, 
$ZT \sim 0.7$ at high temperatures.

A study of the width of the non-equilibrium spectral densities
of both dots, indicates that for small interdot hopping $t$, 
each dot is at the same temperature as the corresponding lead.
Increasing $t$, the effective temperatures of both dots are intermediate between those of the leads and tend to coincide 
for $t$ larger than the resonant level width $\Gamma$.

In this work, we have assumed inversion symmetry. This means 
that the on-site energy of both dots and coupling to the leads are the same.
Asymmetric coupling to the leads can lead to a sizable 
enhancement of the thermoelectric response \cite{asym,mana}. 
In addition, different energies of the dot can lead to important rectification effects \cite{heat}.
It would be interesting to study these asymmetry effects.

\section*{Acknowledgments}

A. A. A. acknowledges financial support provided by PICT 2017-2726 and PICT 2018-01546 of the ANPCyT, Argentina.

\appendix

\section{NONCROSSING APPROXIMATION TREATMENT OF FLUCTUATIONS BETWEEN 1 AND 2 PARTICLES IN THE DQD}

\label{anca} In this appendix we describe in some detail the NCA
approximation of the serial DQD attached to two different conducting leads as
given by Eqs. (\ref{hameff}) and (\ref{hv2}). Performing the explicit
calculations, the matrix elements $D_{\nu \sigma }^{Sm,\xi \sigma ^{\prime
}}=\langle S,m|d_{\nu \sigma }^{\dagger }|\xi \sigma ^{\prime }\rangle $
entering Eq. (\ref{hv2}) to linear order in $\chi =t/(U-U^{\prime })$ turn
out to be

%%%%%%%%%%%%%%%%%%%%%%  R-up %%%%%%%%%%%%%%%%%%%%%%%%%%%%%%%%%%%%%%%%%%%%%%%%%%%%%%%%%%%%%%%%%%%%%%%%
\vskip 0.4cm

\begin{center}
\begin{tabular}{|l||c|c|c|c|}
\hline
$D_{R\uparrow }^{Sm,\xi \sigma ^{\prime }}$ & $e\uparrow $ & 
$o\uparrow $ & $e\downarrow $ & $o\downarrow $ \\ \hline\hline
$1,~~1$ & $\frac{-1}{\sqrt{2}}$ & $\frac{-1}{\sqrt{2}}$ & 0 & 0 \\ \hline
$1,~~0$ & 0 & 0 & $-\frac{1}{2}$ & $-\frac{1}{2}$ \\ \hline
$1,-1$ & 0 & 0 & 0 & 0 \\ \hline
$0,~~0$ & 0 & 0 & $\frac{1}{2}+\chi $ & $\frac{1}{2}-\chi $ \\ \hline
\end{tabular}
\end{center}

\vskip 0.4cm 
%%%%%%%%%%%%%%%%%%%%%%  L-up %%%%%%%%%%%%%%%%%%%%%%%%%%%%%%%%%%%%%%%%%%%%%%%%%%%%%%%%%%%%%%%%%%%%%%%%
%\vskip 0.4cm

\begin{center}
\begin{tabular}{|l||c|c|c|c|}
\hline
$D_{L\uparrow }^{Sm,\xi \sigma ^{\prime }}$ & $e\uparrow $ & 
$o\uparrow $ & $e\downarrow $ & $o\downarrow $ \\ \hline\hline
$1,~~1$ & $\frac{1}{\sqrt{2}}$ & $\frac{-1}{\sqrt{2}}$ & 0 & 0 \\ \hline
$1,~~0$ & 0 & 0 & $\frac{1}{2}$ & $-\frac{1}{2}$ \\ \hline
$1,-1$ & 0 & 0 & 0 & 0 \\ \hline
$0,~~0$ & 0 & 0 & $\frac{1}{2}+\chi $ & $-\frac{1}{2}+\chi $ \\ \hline
\end{tabular}
\end{center}

\vskip 0.4cm

The remaining matrix elements can be obtained from the previous ones
using time-reversal symmetry. The result is

%%%%%%%%%%%%%%%%%%%%%%  R-dn %%%%%%%%%%%%%%%%%%%%%%%%%%%%%%%%%%%%%%%%%%%%%%%%%%%%%%%%%%%%%%%%%%%%%%%%
\vskip 0.4cm

\begin{center}
\begin{tabular}{|l||c|c|c|c|}
\hline
$D_{R\downarrow }^{Sm,\xi \sigma ^{\prime }}$ & $e\uparrow $ & $o\uparrow $
& $e\downarrow $ & $o\downarrow $ \\ \hline\hline
$1,~~1$ & 0 & 0 & 0 & 0 \\ \hline
$1,~~0$ & $-\frac{1}{2}$ & $-\frac{1}{2}$ & 0 & 0 \\ \hline
$1,-1$ & 0 & 0 & $\frac{-1}{\sqrt{2}}$ & $\frac{-1}{\sqrt{2}}$ \\ \hline
$0,~~0$ & $-\frac{1}{2}-\chi $ & $-\frac{1}{2}+\chi $ & 0 & 0 \\ \hline
\end{tabular}
\end{center}

\vskip 0.4cm 
%%%%%%%%%%%%%%%%%%%%%%  L-dn %%%%%%%%%%%%%%%%%%%%%%%%%%%%%%%%%%%%%%%%%%%%%%%%%%%%%%%%%%%%%%%%%%%%%%%%
%\vskip 0.4cm

\begin{center}
\begin{tabular}{|l||c|c|c|c|}
\hline
$D_{L\downarrow }^{Sm,\xi \sigma ^{\prime }}$ & $e\uparrow $ & $o\uparrow $
& $e\downarrow $ & $o\downarrow $ \\ \hline\hline
$1,~~1$ & 0 & 0 & 0 & 0 \\ \hline
$1,~~0$ & $\frac{1}{2}$ & $-\frac{1}{2}$ & 0 & 0 \\ \hline
$1,-1$ & 0 & 0 & $\frac{1}{\sqrt{2}}$ & $\frac{-1}{\sqrt{2}}$ \\ \hline
$0,~~0$ & $-\frac{1}{2}-\chi $ & $\frac{1}{2}-\chi $ & 0 & 0 \\ \hline
\end{tabular}
\end{center}

\vskip 0.4cm

\textit{Auxiliary particle representation}. Now, we introduce auxiliary
bosons and fermions to represent the two-electron and one-electron states, $|S,m\rangle \rightarrow b_{S,m}^{\dagger }|vac\rangle $ and 
$|\xi \sigma \rangle \rightarrow f_{\xi \sigma }^{\dagger }|vac\rangle $ respectively,
where $|vac\rangle $ is a reference state in the auxiliary Fock space. By
using this representation the Hamiltonian in Eq. (\ref{hameff}) reads as
follows

\begin{eqnarray}
H_{eff} &=&\sum_{\xi \sigma }E_{\xi }f_{\xi \sigma }^{\dagger }f_{\xi \sigma
}+\sum_{Sm}E_{S}b_{S,m}^{\dagger }b_{S,m}  \notag  \label{hamauxi} \\
&&+H_{c}+H_{V},
\end{eqnarray}
where the hybridization term becomes

\begin{eqnarray}
H_{V} &=&\sum_{k\nu \sigma }\sum_{\xi \sigma ^{\prime }}\sum_{Sm}V_{k\nu
}\,D_{\nu \sigma }^{Sm,\xi \sigma ^{\prime }}b_{S,m}^{\dagger }f_{\xi \sigma
^{\prime }}c_{\nu k\sigma }  \notag  \label{hv2auxi} \\
&&+\text{H.c.}.
\end{eqnarray}
The auxiliary particles satisfy the constraint $\sum_{\xi \sigma }f_{\xi
\sigma }^{\dagger }f_{\xi \sigma }+\sum_{Sm}b_{S,m}^{\dagger }b_{S,m}=1$,
which represents the completeness relation of the Hilbert space formed by
the states in Eqs. (\ref{states}). Furthermore,
%ACA ECUACIONES CITADAS
%Eq. (\ref{n1-states}) and Eq. (\ref{n2-states}). Furthermore,
the expression of the physical creation operator at each dot is given by 
\begin{equation}
d_{\nu \sigma }^{\dagger }=\sum_{\xi \sigma ^{\prime }}\sum_{Sm}D_{\nu
\sigma }^{Sm,\xi \sigma ^{\prime }}b_{S,m}^{\dagger }f_{\xi \sigma ^{\prime
}}.  \label{phys-operator}
\end{equation}

In this form, the Hamiltonian is suitable for the NCA treatment.

\textit{NonCrossing Approximation}. Starting from the auxiliary Green
functions in the time domain, 
\begin{eqnarray}
G_{\xi \sigma ,\xi ^{\prime }\sigma ^{\prime }}^{>}(t) &=&-i\langle f_{\xi
\sigma }(t)f_{\xi ^{\prime }\sigma ^{\prime }}^{\dagger }(0)\rangle  \notag
\label{aux-green-definition} \\
G_{\xi \sigma ,\xi ^{\prime }\sigma ^{\prime }}^{<}(t) &=&+i\langle f_{\xi
^{\prime }\sigma ^{\prime }}^{\dagger }(0)f_{\xi \sigma }(t)\rangle  \notag
\\
B_{Sm}^{>}(t) &=&-i\langle b_{S,m}(t)b_{S,m}^{\dagger }(0)\rangle  \notag \\
B_{Sm}^{<}(t) &=&-i\langle b_{S,m}^{\dagger }(0)b_{S,m}(t)\rangle ,
\label{gfne}
\end{eqnarray}
and from Eq. (\ref{hamauxi})  we can write down the auxiliary
selfenergies for the auxiliary particles as follows

\begin{widetext}
\begin{eqnarray}\label{sigmas}
\Sigma^{\gtrless}_{\xi\sigma',\xi'\sigma''}(\omega)&=& \pm\sum_{\nu\sigma,Sm_{S}}\Gamma_{\nu\sigma}D_{\nu\sigma}^{S m_{S},\xi \sigma'}D_{\nu\sigma}^{\xi'\sigma'',S m_{S}}\int \frac{d\omega'}{2\pi}f_{\nu}(\pm\omega')B^{\gtrless}_{S m_{S}}(\omega'+\omega+\mu_{\nu})\nonumber\\
\Pi^{\gtrless}_{S m_{S}}(\omega)&=& \pm\sum_{\nu\sigma,\xi\sigma',\xi'\sigma''}\Gamma_{\nu\sigma}D_{\nu\sigma}^{S m_{S},\xi \sigma'}D_{\nu\sigma}^{\xi'\sigma'',S m_{S}}\int \frac{d\omega'}{2\pi}f_{\nu}(\pm\omega')G^{\gtrless}_{\xi\sigma',\xi'\sigma''}(\omega'+\omega-\mu_{\nu})
\end{eqnarray}
\end{widetext}where 
$f_{\nu }(\omega )=[1+e^{\omega /T_{\nu }}]^{-1}$ is the
Fermi function at temperature $T_{\nu }$.

The complete set of lesser and greater selfenergies are purely imaginary
functions. Therefore the imaginary part of the retarded 
selfenergies are given by 
\begin{eqnarray}
\mbox{Im}~\Sigma _{\xi \xi ^{\prime },\sigma }^{r}(\omega ) &=&\Sigma _{\xi
\xi ^{\prime },\sigma }^{>}(\omega )/2  \notag \\
\mbox{Im}~\Pi _{Sm_{S}}^{r}(\omega ) &=&\Pi _{Sm_{S}}^{>}(\omega )/2,
\label{sigmar}
\end{eqnarray}
while the real parts can be obtained from the imaginary ones by using a
Kramers-Kronig transformation.

In addition, 
$G^{\lessgtr}_{\xi\xi} = i\mathbb{G}^{\lessgtr}_{\xi\xi}$, 
$G^{\lessgtr}_{\xi\xi^{\prime }} + G^{\lessgtr}_{\xi^{\prime }\xi} = i \big( \mathbb{G}^{\lessgtr}_{\xi\xi^{\prime }} + 
\mathbb{G}^{\lessgtr}_{\xi^{\prime }\xi}\big)$, with $\xi\neq\xi^{\prime }$, and $B^{\lessgtr} = i \mathbb{B}^{\lessgtr} $ 
are purely imaginary. Regarding signs, 
$\mathbb{G}^{<}_{\xi\xi} > 0$ and $\mathbb{G}^{>}_{\xi\xi} < 0 $ and 
$\mathbb{B}^{\lessgtr} < 0 $ (at least in equilibrium). 
The constraint, being
a positive magnitude, reads $Q=\int \frac{d\omega}{2\pi} \big( %
\sum_{\xi,\sigma}\mathbb{G}^{<}_{\xi\xi,\sigma}(\omega) -\sum_{S m_{S}}%
\mathbb{B}^{<}_{S m_{S}}(\omega)\big)$.

%both $\Sigma^{>}_{\xi\xi}$ (diagonal) and $\Pi^{>}$ should be $>0$. Therefore  

For the fermion retarded Green functions we obtain

\begin{eqnarray}  \label{green-matrix-fermions}
G^{r}_{ee,\sigma}(\omega)&=&(\omega-E_{o}-\Sigma^{r}_{oo,\sigma}(\omega))/\mathbb{D}(\omega)  \notag \\
G^{r}_{oo,\sigma}(\omega)&=&(\omega-E_{e}-\Sigma^{r}_{ee,\sigma}(\omega))/\mathbb{D}(\omega)  \notag \\
G^{r}_{eo,\sigma}(\omega)&=&\Sigma^{r}_{eo,\sigma}(\omega)/\mathbb{D}(\omega)
\notag \\
G^{r}_{oe,\sigma}(\omega)&=&\Sigma^{r}_{oe,\sigma}(\omega)/\mathbb{D}(\omega)
\notag \\
\mathbb{D}(\omega)&=&(\omega-E_{o}-\Sigma^{r}_{oo,\sigma}(\omega))
(\omega-E_{e}-\Sigma^{r}_{ee,\sigma}(\omega))  \notag \\
&&- \Sigma^{r}_{eo,\sigma}(\omega)\Sigma^{r}_{oe,\sigma}(\omega).
\end{eqnarray}

On the other hand, the bosons retarded Green functions read as follows 
\begin{eqnarray}
B^{r}_{S m_{S}}(\omega)&=&\frac{1}{\omega-E_{S}-\Pi^{r}_{S m_{S}}(\omega)}.
\end{eqnarray}

The selfconsistent procedure closes with the matrix relations

\begin{eqnarray}
G^{\gtrless}&=& G^{r} \Sigma^{\gtrless} G^{a}  \notag \\
B^{\gtrless}&=& B^{r} \Pi^{\gtrless} B^{a}.
\end{eqnarray}

An example for a particular matrix element is 
\begin{equation}
G_{eo}^{>}=\sum_{\mu \nu }G_{e\mu }^{r}\Sigma _{\mu \nu }^{>}G_{\nu o}^{a}
\end{equation}

\textit{Physical Green functions}. By using the auxiliary expression of the
physical operator in Eq. (\ref{phys-operator}) the physical Green functions
for each dot are given by 
\begin{widetext}
\begin{eqnarray}\label{green-fis}
 G^{\gtrless}_{\nu\nu',\sigma}(\omega)&=&-\frac{i}{Q}\sum_{\xi\xi'\sigma'}\sum_{S m_{S}}
 D_{\nu\sigma}^{\xi\sigma',S m_{S}} D_{\nu'\sigma}^{S m_{S},\xi'\sigma'}
 \int \frac{d\omega'}{2\pi} G^{\lessgtr}_{\xi\xi',\sigma'}(\omega')
 B^{\gtrless}_{S m_{S}}(\omega'+\omega)
\end{eqnarray}
\end{widetext}

\section{AN ALTERNATIVE EXPRESSION FOR THE CHARGE CURRENT}

\label{current-delta}

In this appendix we give an alternative expression for
the calculation of the charge current for the Hamiltonian Eqs. (1) to (4) of
the main text. Current conservation along the system leads to the following
relation, 
\begin{equation}
\frac{d}{dt}\Big(N_{L}+n_{L}\Big)=-\frac{d}{dt}\Big(N_{R}+n_{R}\Big),
\end{equation}
where $n_{i}=\sum_{\sigma }n_{i\sigma }$ and $N_{i}=\sum_{k_{i}\sigma
}c_{k_{i}\sigma }^{\dagger }c_{k_{i}\sigma }$. Therefore, the electric
current can be computed from the left part of the system by using the
Heisenberg equation of motion 
\begin{equation}
\hat{J}_{L}^{C}\equiv \frac{e}{\hslash }\frac{d}{dt}\Big(N_{L}+n_{L}\Big)=-i \frac{e}{\hslash }\Big[N_{L}+n_{L},H\Big].
\end{equation}
After some algebra and by using the identity $[n_{i\sigma },n_{j\sigma
^{\prime }}]=-\delta _{ij}\delta _{\sigma \sigma ^{\prime }}(c_{j\sigma
^{\prime }}^{\dagger }c_{i\sigma }+c_{i\sigma }^{\dagger }c_{j\sigma
^{\prime }})$ the right hand side simplifies to 
\begin{equation}
\hat{J}_{L}^{C}=-\frac{ie}{\hslash }t\sum_{\sigma }\Big(d_{R\sigma
}^{\dagger }d_{L\sigma }-d_{L\sigma }^{\dagger }d_{R\sigma }\Big),
\end{equation}
which is nothing but the usual form of the current operator between two
sites of a linear chain with a hopping equal to $t$. The Fourier
transform $G_{ij,\sigma}^{<}(\omega )$ of the time-dependent Green's functions 
$G_{ij,\sigma}^{<}(t)=
i \langle d_{j\sigma}^{\dagger }d_{i\sigma}(t) \rangle$ 
provides the average of 
$\hat{J}_{L}^{C}$ leading the expression for the current in the form 
\begin{eqnarray}
J_{L}^{C} &=&\frac{e}{\hslash }t\sum_{\sigma }
\int \frac{d\omega }{2\pi } \big(G_{LR,\sigma}^{<}(\omega )-G_{RL,\sigma}^{<}(\omega )\big)  \label{current_3} \notag \\
&=&\frac{e}{\hslash }t\sum_{\sigma }\int \frac{d\omega }{2\pi }
\big(G_{eo,\sigma}^{<}(\omega )-G_{oe,\sigma}^{<}(\omega )\big). 
\end{eqnarray}
Note that the last expression follows from $\langle d_{R\sigma }^{\dagger
}d_{L\sigma }-d_{L\sigma }^{\dagger }d_{R\sigma }\rangle =\langle d_{e\sigma
}^{\dagger }d_{o\sigma }-d_{o\sigma }^{\dagger }d_{e\sigma }\rangle $. Using
that $G_{oe,\sigma}^{<}(\omega )$ is the complex conjugate of \ $G_{eo,\sigma}^{<}(\omega )
$, being both purely imaginary, and adding both spins, one arrives at Eq. (\ref{currd}) of the main text.

\bibliography{references}

\end{document}